\documentclass[aps,prl,showpacs,twocolumn,groupedaddress]{revtex4}

\usepackage{graphicx}
\usepackage{dcolumn}
\usepackage{bm}
\usepackage{color}

\begin{document}


\title{Magnetization Plateaus in the Spin-1/2 Kagome Antiferromagnets: \\ Volborthite and Vesignieite}

%

\author{Y. Okamoto$^*$, M. Tokunaga, H. Yoshida$^{\dag}$, A. Matsuo, K. Kindo, and Z. Hiroi}
\affiliation{
Institute for Solid State Physics, University of Tokyo, Kashiwanoha 5-1-5, Kashiwa 277-8581, Japan\
}

\date{\today}

\begin{abstract}
The magnetization of two spin-1/2 kagome antiferromagnets, volborthite and vesignieite, has been measured in pulsed magnetic fields up to 68 T. A magnetization plateau is observed for each compound near the highest magnetic field. Magnetizations at saturation are approximately equal to 0.40$M_{\mathrm{s}}$ for both compounds, where $M_{\mathrm{s}}$ is the fully saturated magnetization, irrespective of a difference in the distortion of the kagome lattice between the two compounds. 
It should be noted that these values of magnetizations are significantly larger than $M_{\mathrm{s}}$/3 predicted theoretically for the one-third magnetization plateau in the spin-1/2 kagome antiferromagnet. The excess magnetization over $M_{\mathrm{s}}$/3 is nearly equal to the sum of the magnetizations gained at the second and third magnetization steps in volborthite, suggesting that there is a common origin for the excess magnetization and the magnetization steps.
\end{abstract}

\pacs{Valid PACS appear here}
\maketitle

Various nontrivial spin states can appear in geometrically frustrated antiferromagnets~\cite{1}. Before full polarization at sufficiently large magnetic fields, a specific spin arrangement that is compatible with the lattice geometry can be stabilized in a range of magnetic fields. This state manifests itself in the $M$-$H$ curve as a plateau at a simple fractional value of the fully saturated magnetization $M_{\mathrm{s}}$, known as the magnetization plateau~\cite{2}. 

The most intensively studied examples of the magnetization plateau are the classical spin antiferromagnets on a triangular lattice. Their ground state at zero magnetic field is the 120 degree structure, which is a coplanar state with a $\sqrt{3} \times \sqrt{3}$ superlattice. In sufficiently large magnetic fields, a magnetization plateau should appear at one third of $M_{\mathrm{s}}$ in the case of Heisenberg spins with a finite Ising anisotropy~\cite{3}. This state is characterized by a collinear spin arrangement with up-up-down (uud) spins for each triangle. Model compounds studied experimentally so far include 
RbFe(MoO$_4$)$_2$, CsFe(SO$_4$)$_2$~\cite{4}, EuC$_6$~\cite{5} and GdPd$_2$Al$_3$~\cite{6}.
All of these examples exhibit 1/3 magnetization plateaus, which are probably stabilized by Ising anisotropy.

On the other hand, it is theoretically predicted for quantum-spin antiferromagnets on the triangular lattice that a 1/3 magnetization plateau appears, owing to quantum fluctuations even in a pure Heisenberg spin system free from Ising anisotropy~\cite{2,8}. A possible spin arrangement at the plateau is characterized by quantum-mechanical superpositions of the uud spin configuration for each triangle, which has been selected by the order-by-disorder mechanism. Experimentally, a 1/3 magnetization plateau has been found only in the spin-1/2 antiferromagnet Cs$_2$CuBr$_4$, which contains a distorted triangular lattice made up of Cu$^{2+}$ spins~\cite{9}. 

In the case of the kagome lattice, which is more frustrated than the triangular lattice, the ground state is expected to be a spin liquid with or without a spin gap~\cite{SL1,SL2}. In magnetic fields, a similar 1/3 magnetization plateau 
may appear, even in a pure Heisenberg spin-1/2 antiferromagnet, as suggested by exact diagonalization studies on finite-size clusters~\cite{11,12}. The plateau appears above $H_{\mathrm{p}1} \sim$ 0.9$J$,
where $J$ is the nearest-neighbor exchange coupling constant. 

Several compounds are known to be candidates for the kagome antiferromagnet (KAFM). Recently, copper minerals, such as herbertsmithite~\cite{13}, volborthite~\cite{14} and vesignieite~\cite{15}, have attracted much attention as candidates for the spin-1/2 KAFM. However, there has been no experimental evidence for the magnetization plateau in any kagome compound so far. In the case of herbertsmithite, it would be experimentally difficult to reach the 1/3 magnetization plateau because of a large $J$ value of 170 K~\cite{16}. As a consequence, inaccessibly high magnetic fields above 200 T are needed to detect the 1/3 plateau according to the theoretical expectation of $H_{\mathrm{p}1} \sim$ 0.9$J$. In contrast, there is a chance to observe a 1/3 plateau state experimentally in volborthite and vesignieite, which have relatively small $J$ values of 77 and 55 K~\cite{15,17}, respectively. Here we report state-of-the-art high-magnetic-field magnetization measurements up to 68 T. We have found for the first time in the KAFMs a saturation of the magnetization toward a plateau for both compounds, which, surprisingly however, occurs at magnetizations of $\sim$0.40$M_{\mathrm{s}}$, significantly larger than $M_{\mathrm{s}}$/3.

Volborthite Cu$_3$V$_2$O$_7$(OH)$_2 \cdot$2H$_2$O and vesignieite BaCu$_3$V$_2$O$_8$(OH)$_2$ comprise Cu$^{2+}$ ions carrying spin 1/2 on kagome lattices~\cite{14,15}. The kagome lattice of the former is slightly distorted, while the latter is almost isotropic. However, the nature of the magnetic couplings in volborthite still remain controversial. Recent density-functional-theory calculations claim that the kagome lattice consists of the frustrated $J_1$-$J_2$ chains together with the third spins in between and thus can be far from the anisotropic kagome model~\cite{18}. Anyway, the advantage of volborthite over other compounds is that one can prepare a high quality sample containing fewer impurity spins, \textit{i}.\textit{e}., only 0.07\% of the total spin. This allows one to investigate the intrinsic properties of the compound at low temperatures~\cite{17}. The magnetic susceptibility shows no anomaly, indicating long range magnetic ordering down to 60 mK, and approaches a large finite value at $T$ = 0, which provides evidence for a gapless, liquid-like ground state. In contrast, $^{51}$V-NMR measurements reveal a magnetic transition at 1 K to a peculiar phase that is characterized by the presence of dense low energy excitations and unusually slow spin dynamics~\cite{19}. These results strongly suggest that the ground state of volborthite is not a simple long-range order but something else. Moreover, three magnetization steps are observed at magnetic fields of 4.3 T, 25.5 T and 46 T in magnetization measurements up to 55 T~\cite{17}. 
On the other hand, the magnetization plateau has not been observed in the previous study up to 55 T, which corresponds to 0.5$J$.

Vesignieite is certainly a good candidate for the KAFM but suffers from low sample quality as herbertsmithite~\cite{20}, typically containing several \% of impurity spins that may mask intrinsic properties, particularly at low temperatures. Intrinsic magnetic susceptibility obtained by subtracting the contribution of impurity spins exhibits neither long range order nor a spin-glass transition down to 2 K and goes to a large finite value toward $T$ = 0~\cite{15}. Hence, the ground state of vesignieite may be gapless, as in volborthite. High-field magnetization of vesignieite has not yet been studied. We expect that the smaller $J$ in vesignieite allows us to study magnetization over a wider range of $H$/$J$ than in volborthite. In addition, information on the effects of distortion of or deviations from the kagome lattice may be deduced by making comparisons of the magnetization process between the two compounds. 

High-magnetic-field magnetization measurements on powder samples of volborthite and vesignieite were performed by the induction method using a multilayered pulsed magnet up to 68 T.
The time evolution of magnetization was recorded on increaseing and decreasing magnetic field
in a duration time of 6 and 36 msec for 68 and 55 T data, respectively.
Since it was difficult to obtain absolute values of magnetization by this method, we have corrected the data to fit another magnetization curve measured on the same sample up to 7 T in a commercial SQUID magnetometer (MPMS, Quantum Design). 
A good linear response in magnetization up to the maximum fields and thus
a reliability in the magnetization values have been confirmed in various 
magnetic compounds~\cite{HF}.
Powder samples of volborthite and vesignieite were prepared by the hydrothermal method~\cite{15,17}. The volborthite sample is the same as that used in the previous magnetization measurements up to 55 T~\cite{17}, while the vesignieite sample has a similar quality in terms of impurity content as previous ones used in the magnetic susceptibility measurements~\cite{15}. 
No tendency for the preferred orientation of the powder samples was detected.

\begin{figure}
\includegraphics[width=8.5cm]{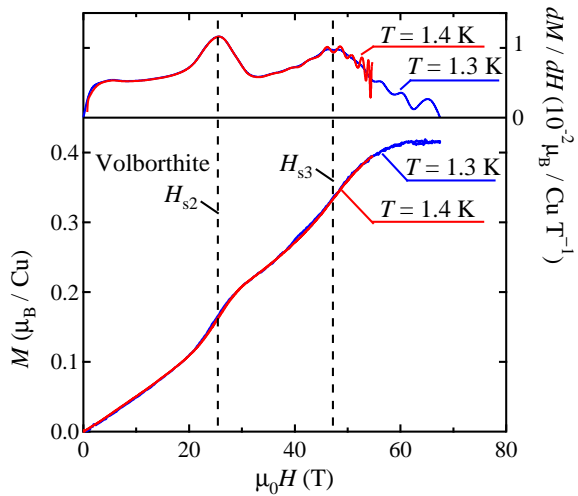}
\caption{\label{fig1} (Color online) Comparison of two magnetization curves for a powder sample of volborthite measured previously up to 55 T at 1.4 K~\cite{17} and up to 68 T at 1.3 K in the present study. Each dataset includes two curves measured on increasing and decreasing magnetic fields, which overlap to each other completely. First, second, and third step fields, $H_{\mathrm{s}1}$, $H_{\mathrm{s}2}$, and $H_{\mathrm{s}3}$, are indicated as broken lines. The derivative of each $M$-$H$ curve is shown at the top.
}

\end{figure}

Figure 1 shows a $M$-$H$ curve for volborthite measured up to 68 T at 1.3 K, which is compared with the previous data collected up to 55 T at 1.4 K~\cite{17}. The fact that the two curves exactly overlap demonstrates good experimental reproducibility even at such high fields.
The second and third magnetization steps are clearly observed in both data at $\mu_0H_{\mathrm{s}2}$ = 25.6 T and $\mu_0H_{\mathrm{s}3}$ = 47 T. These steps are defined at the maxima of the derivative curves shown in the top of Fig. 1. Magnetizations at the second and third steps are $M_{\mathrm{s}2}$ = 0.168 $\mu_{\mathrm{B}}$/Cu and $M_{\mathrm{s}3}$ = 0.33 $\mu_{\mathrm{B}}$/Cu. They correspond to 0.156 and 0.31$M_{\mathrm{s}}$, respectively, providing $M_{\mathrm{s}}$ given by $gS\mu_{\mathrm{B}}$, where $g$ is the Lande g-factor. 
We take the powder averaged value of 2.15 from ESR measurements~\cite{21}; no preferred orientation occurred even in high magnetic fields.
The first magnetization step is present at $\mu_0H_{\mathrm{s}1}$ = 4.3 T and $M_{\mathrm{s}1}$ = 0.019 $\mu_{\mathrm{B}}$/Cu but too small to observe in Fig. 1. 

In the 68 T curve, magnetization shows distinct saturation behavior and becomes nearly constant at 0.42 $\mu_{\mathrm{B}}$/Cu above $\sim$60 T, which is reminiscent of the magnetization plateau reported in various other spin systems. This magnetization corresponds to 0.39$M_{\mathrm{s}}$, which is significantly larger than $M_{\mathrm{s}}$/3. Note that $M_{\mathrm{s}}$/3 has been already exceeded at around 50 T in the $M$-$H$ curves. Obviously, the excess magnetization of $M_{\mathrm{ex}}$ = 0.06$M_{\mathrm{s}}$ over $M_{\mathrm{s}}$/3 is much larger than the experimental ambiguity.

\begin{figure}
\includegraphics[width=8cm]{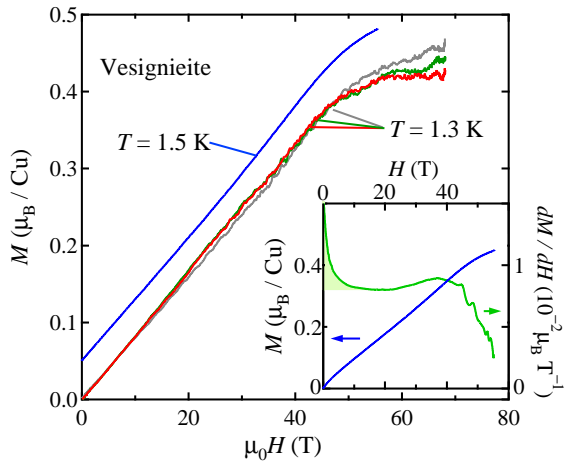}
\caption{\label{fig2} (Color online) $M$-$H$ curves of a powder sample of vesignieite. 
Measurements were performed once up to 55 T at 1.5 K and three times up to 68 T at 1.3 K. 
For clarity, the former curve is shifted upward by 0.05 $\mu_{\mathrm{B}}$/Cu. 
The data after the removal of impurity contributions are shown.
The inset shows an $M$-$H$ curve up to 55 T at 1.5 K before the removal of impurity contributions
and its field-derivative curve.
The hatched region in the derivative curve below 10 T represents an additional contribution from 1.5\% impurity spins.
}

\end{figure}

The magnetization curves for vesignieite up to 55 and 68 T are shown in Fig. 2. The contributions from the 1.5\% impurity spins have been already removed, which caused a reduction in magnetization by $\sim$0.01$M_{\mathrm{s}}$~\cite{imp}.
A 55 T curve taken at 1.5 K is smooth, while three sets of 68 T data at 1.3 K are noisy and deviate considerably from each other above 60 T because of the experimental difficulties involved in using a high-field pulsed magnet.
However, it is highly probable that there is a clear tendency for saturation above 50 T, which is already discernible in the 55 T curve and more readily observed in the 68 T curves. The magnetization at the plateau is $\sim$0.43 $\mu_{\mathrm{B}}$/Cu, calculated by averaging the values at 60 T, which corresponds to 0.40$M_{\mathrm{s}}$ using $g$ = 2.14 from ESR measurements~\cite{24}. Remarkably, this plateau magnetization is much larger than $M_{\mathrm{s}}$/3 and almost equal to that observed in volborthite.

It has been clearly demonstrated in the present study that both volborthite and vesignieite exhibit saturating behavior in the $M$-$H$ curves at high magnetic fields, as expected theoretically for the spin-1/2 KAFM. Figure 3 compares their $M$-$H$ curves normalized by $g$ and $J$ ($J_{\mathrm{av}}$); 2.15 and 77 K for volborthite, and 2.14 and 55 K for vesignieite~\cite{15,17,21,24}. Note that two saturation values occur at nearly equal magnetizations of $\sim$0.40$M_{\mathrm{s}}$ 
or $\sim$(2/5)$M_{\mathrm{s}}$,
which is $\sim$20\% larger than $M_{\mathrm{s}}$/3. This deviation is unlikely to be due to spatial anisotropy of the kagome lattice and is probably intrinsic for the spin-1/2 KAFM. On the other hand, the lower critical fields are roughly 0.5$J$ and 0.7$J$ for volborthite and vesignieite, respectively. The smaller $H_{\mathrm{p}1}$ of volborthite may be caused by the larger spatial anisotropy, because an exact diagonalization study suggests that $H_{\mathrm{p}1}$ decreases with increasing anisotropy~\cite{25}. Therefore, serious discrepancies regarding the magnetization plateau exist between experiments on volborthite and vesignieite and theory for the spin-1/2 KAFM.

\begin{figure}
\includegraphics[width=7.5cm]{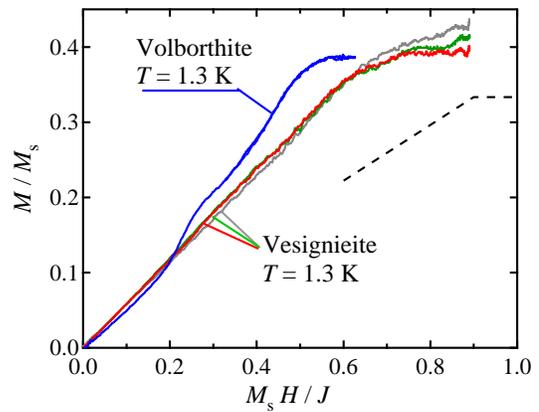}
\caption{\label{fig3} (Color online) Normalized magnetization curves for volborthite and vesignieite. The 68 T data have been translated by using the ($g$, $J$/$k_{\mathrm{B}}$) = (2.15, 77 K) for volborthite~\cite{17,21} and (2.14, 55 K) for vesignieite~\cite{15,24}. The broken line represents a theoretical curve for the Heisenberg spin-1/2 KAFM obtained by exact diagonalization study~\cite{11}. 
}

\end{figure}

The large deviation of the plateau magnetization from $M_{\mathrm{s}}$/3 observed in volborthite and vesignieite is rather surprising, because all of magnetization plateaus so far observed in other frustrated magnets on triangle-based lattices occur precisely at $M_{\mathrm{s}}$/3. This is robust and independent of whether the system consists of classical or quantum spins, Ising or Heisenberg spins, and a distorted or undistorted lattice~\cite{4,5,6,9}. 
Dzyaloshinskii-Moriya interactions may also have little effect on the magnetization value 
at the plateau~\cite{DM}. Hence, it is quite unusual, and there must be some specific mechanism to enhance the magnetization, which may be unique for the KAFM. The 1/3 plateau state is commonly based on the uud spin configulation on one triangle. Thus, in order to explain the excess magnetization, one has to assume a more expanded object on the kagome net. 

We point out here a possible relation between the excess magnetization $M_{\mathrm{ex}}$ over $M_{\mathrm{s}}$/3 and the magnetization steps observed in volborthite. $M$ is often proportional to $H$ for a magnetic state without a ferromagnetic component. In fact, this is the case for $M$ in phase II, $M$(II), as well as that in phase III, $M$(III), as shown in Fig. 4. It is considered that $M$ gains a certain amount at each step, in addition to the $M$ below. Ignoring a tiny jump at the first step, the $M$(II) is estimated as 0.48$M_{\mathrm{s}}^2 H$/$J$ from a linear fit below the first step. Similarly, $M$(III) and $M$(IV) are determined as 0.69$M_{\mathrm{s}}^2 H$/$J$ and 0.73$M_{\mathrm{s}}^2 H$/$J$, respectively. Then, the jumps at the second and third steps, $\Delta M_{\mathrm{s}2}$ and $\Delta M_{\mathrm{s}3}$, are $\Delta M_{\mathrm{s}2}$ = (0.69 $-$ 0.48)$M_{\mathrm{s}}^2 H_{\mathrm{s}2}$/$J$ = 0.050$M_{\mathrm{s}}$ and $\Delta M_{\mathrm{s}3}$ = (0.73 $-$ 0.69)$M_{\mathrm{s}}^2 H_{\mathrm{s}3}$/$J$ = 0.017$M_{\mathrm{s}}$. Interestingly enough, the summation of $\Delta M_{\mathrm{s}2}$ and $\Delta M_{\mathrm{s}3}$ reaches 0.067$M_{\mathrm{s}}$, which is close to $M_{\mathrm{ex}}$ = 0.06$M_{\mathrm{s}}$. This strongly suggests that there is a common origin for the two phenomena.

\begin{figure}
\includegraphics[width=8.5cm]{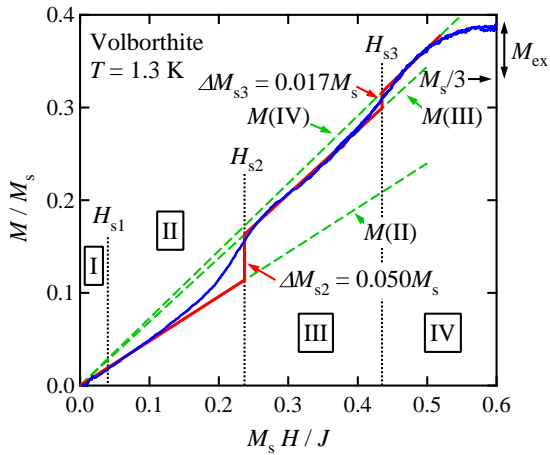}
\caption{\label{fig4} (Color online) Normalized $M$-$H$ curve of volborthite obtained at 1.3 K. First, second, and third step fields, $H_{\mathrm{s}1}$, $H_{\mathrm{s}2}$, and $H_{\mathrm{s}3}$, are shown by vertical dotted lines. Magnetic phases between them are denoted by I, II, III, and IV. A linear magnetization for each phase is shown by a dashed line. Excess magnetization over $M_{\mathrm{s}}$/3, $M_{\mathrm{ex}}$, is also indicated. 
}

\end{figure}

It is likely that similar magnetization steps also exist in vesignieite. A careful observer finds in the $M$-$H$ curves of vesignieite in Fig. 2 small upward deviations from a linear contribution, one at $\sim$30 T in the 55 T data, and two at $\sim$15 and $\sim$35 T in the 68 T data. Since they are much smaller than the second step in volborthite, it is difficult to conclude the presence of magnetization steps in vesignieite. We think that the reason is the poor sample quality of vesignieite in comparison with volborthite. The 1.5\% impurity spins included in the vesignieite sample is much larger than the 0.07\% in volborthite and may mask the intrinsic magnetization or seriously disturb the ground state. 
One might observe similar magnetization steps in vesignieite with a higher quality sample. 
We believe that common physics underlies in the two compounds, which represents the intrinsic nature of spin-1/2 KAFM, or at least that of distorted KAFMs, irrespective of the magnitude of distortion.

In conclusion, saturating behavior in magnetization toward the magnetization plateau has been found for two spin-1/2 KAFMs, volborthite and vesignieite, which comprise a distorted and an almost isotropic kagome lattice, respectively. The plateaus appear at nearly equal magnetizations $\sim$0.40$M_{\mathrm{s}}$ or close to $(2/5)M_{\mathrm{s}}$ which are $\sim$20\% larger than $M_{\mathrm{s}}$/3 expected for the 1/3 magnetization plateau in the spin-1/2 KAFM. The deviation from the fractional magnetization may be related to the magnetization steps observed in volborthite and possibly also present in vesignieite. We believe that to uncover the mysteries on the kagome lattice would lead us to a further understanding of the frustration physics. 

This work was partly supported by a Grant-in-Aid for Scientific Research on Priority Areas ``Novel States of Matter Induced by Frustration'' (19052003).


\begin{thebibliography}{199}
\bibitem{imp}
To obtain intrinsic magnetization $M_{\mathrm{int}}$ for vesignieite, we have subtracted the contribution of impurity spins, $M_{\mathrm{imp}}$, from the experimental data, $M_{\mathrm{exp}}$: $M_{\mathrm{int}}$ = $M_{\mathrm{exp}} - M_{\mathrm{imp}}$. As shown in the inset of Fig. 2, $M_{\mathrm{exp}}$ shows convex upward curvature below 10 T and then increases linearly, which is clearly represented in the differential curve. The initial convex-upward contribution at low fields must come from impurity spins having weak magnetic interactions with surrounding spins. Provided that $M_{\mathrm{int}}$ $\propto$ $H$ below 10 T with the same slope as in the higher-field region, $dM_{\mathrm{int}}$/$dH$ is estimated to be 8.0 $\times$ 10$^{-3}$ $\mu_{\mathrm{B}}$ T$^{-1}$, as shown in the inset of Fig. 2. Then, $dM_{\mathrm{imp}}$/$dH$ is obtained by subtracting this constant component from $dM_{\mathrm{exp}}$/$dH$. By integrating $dM_{\mathrm{imp}}$/$dH$ over $H$, $M_{\mathrm{imp}}$ is estimated to be 0.017 $\mu_{\mathrm{B}}$/Cu, corresponding to 1.5\% of the total Cu spins. $M_{\mathrm{int}}$ is obtained by removing the $M_{\mathrm{imp}}$ from the $M_{\mathrm{exp}}$ and plotted in the main panel of Fig. 2.
\end{thebibliography}
\end{document}